\documentclass[prb,twocolumn,showpacs,preprintnumbers,amsmath,amssymb]{revtex4}
\usepackage{graphicx}  
\begin{document}
\preprint{}

\title{Quantum and classical localisation, the spin quantum Hall effect and
generalisations}
\author{E.~J.~Beamond$^1$, John Cardy$^{1,2}$, and J.~T.~Chalker$^1$}
\affiliation{$^1$Theoretical Physics, University of Oxford, 1 Keble Road,
Oxford  OX1 3NP, UK\\
$^2$All Souls College, Oxford}
\date{\today}

\begin{abstract}
We consider network models for localisation problems belonging
to symmetry class C. This symmetry class arises in a description of
the dynamics of quasiparticles for disordered spin-singlet superconductors
which have a Bogoliubov -- de Gennes Hamiltonian that is 
invariant under spin rotations but not under time-reversal.
Our models include but also generalise the one studied previously
in the context of the spin quantum Hall effect. For these
systems we express the disorder-averaged conductance and
density of states in terms of sums over certain classical random
walks, which are self-avoiding and have attractive interactions.
A transition between localised and extended phases of the
quantum system maps in this way to a similar transition for the
classical walks. In the case of the spin quantum Hall
effect, the classical walks are the hulls of percolation clusters,
and our approach provides an alternative derivation of a
mapping first established by Gruzberg, Read and Ludwig, 
Phys. Rev. Lett. {\bf 82}, 4254 (1999).
\end{abstract}

\pacs{72.15.Rn, 05.40.Fb, 64.60.Ak, 05.50.+q}
\maketitle

\section{Introduction}

Localisation of a particle moving in a random environment
may occur both quantum-mechanically and with classical dynamics,
but the phenomenon is very different in the two cases.
In this paper we discuss a class of quantum-mechanical localisation
problems for which some physical quantities can be expressed 
exactly in terms of averages taken in a classical counterpart.
The equivalence holds despite the fact that interference
effects dominate the behaviour of the quantum systems.

Disordered quantum systems can in general be classified according to their
symmetries under time reversal and spin rotation. Three such symmetry
classes are represented by the Wigner-Dyson random matrix ensembles,
while an additional seven have been identified more recently.
The models we study here belong to one of
these additional classes, termed class C by Altland and Zirnbauer.\cite{alt-zirn} One feature which distinguishes
systems belonging
to each of the additional symmetry classes 
from those in the Wigner-Dyson classes is that 
they have a special energy in their 
spectrum, with eigenstates occurring in pairs either side of this energy.
Some of the additional classes have realisations
as Bogoliubov -- de Gennes Hamiltonians for quasiparticles in disordered
superconductors, where pairing interactions are treated at the mean-field level.
Here, the special energy is the chemical potential in the superconductor
and eigenstates are related in pairs by a particle-hole transformation.
In particular, class C arises for quasiparticles in a 
spin-singlet superconductor in which time-reversal symmetry is broken
for orbital motion but Zeeman splitting is negligible.\cite{alt-zirn}
Since quasiparticle charge is not conserved in a superconductor, 
experiments to investigate quasiparticle dynamics in these systems 
must probe thermal or spin transport. 
Moreover, as the characteristic
features of the symmetry class appear only close to the chemical potential,
it is particularly gapless superconductors that are interesting:
cuprate superconductors in the mixed state constitute a conspicuous
example.

The models we study are especially simple realisations of their symmetry
class. They are obtained as generalisations of the network model originally
introduced to describe localisation in the context of the integer 
quantum Hall plateau transition.\cite{chalk-codd} Thus, they are formulated
in the language of scattering theory and represent quantum particles,
in general with $N$-component wavefunctions, 
propagating on the directed links, or edges, 
of a lattice and scattering between links
at nodes. 
The symmetry of class C restricts $N$ to even values, while our approach
requires that all nodes of the lattice have two ingoing
and two outcoming links.
For these models, we are concerned with the density of states,
obtained from the time-evolution operator,
and with the disorder-averaged conductance of a finite sample,
calculated from the Landauer formula.
In both cases, our starting point is an expansion for the Green
function as a sum over Feynman paths. Our central result is that 
the terms in this sum which survive after disorder-averaging
can be interpreted as self-avoiding classical random walks 
with attractive, short-range interactions.

A particular network model from class C,
in two dimensions and with $N=2$,
has been studied previously.\cite{kag-prb,kag-prl,grl,smf,cardy-sqhe}
It shows the so-called spin quantum Hall effect, having 
two insulating phases, with quantised 
values of Hall conductance differing by an
integer, separated in the phase diagram by
a delocalisation transition which is analogous
to the quantum Hall plateau transition.
In a remarkable paper, 
using supersymmetry to perform disorder averages,
Gruzberg, Read and Ludwig\cite{grl} (GRL) have shown that many physical 
quantities of interest in this model can be determined from the
properties of the perimeters, or hulls, of classical percolation clusters 
in two dimensions. The approach we describe here provides 
an alternative derivation of their results, using more
elementary, non-supersymmetric methods, as well as an extension to
other lattices, including ones in more than two dimensions and
irregular lattices for which transfer matrix methods are inappropriate. 
It also extends to any even integer $N$. 
Our expressions give disorder-averaged physical quantities 
for the quantum system in terms
of averages over classical random walks
on the same lattice. In the case treated by GRL
these walks are simply percolation hulls, for which
many analytical results are available. By contrast, in
the general case the properties of the classical walks are not known.
Nevertheless, the classical problem is much simpler
than the original quantum problem, and we are able to construct 
further examples for which it is tractable.
We remark that a different type of connection between quantum Hall plateau
transitions and percolation, based on the classical limit,
has been discussed recently in Refs.~\onlinecite{gurarie}
and \onlinecite{moore}.

There are some important qualitative differences between the properties
of systems from Wigner-Dyson classes and those from
the additional symmetry classes. In particular, while single-particle 
quantities such as the density of states are smooth functions
of energy in the former case, in the latter case they may have singularities 
at the special energy, which we take to be zero in the
following. 
This is illustrated by previous results 
on behaviour of models from class C, obtained using 
a variety of techniques.\cite{zirn-su2,sigma1,sigma,sent-fish,dmpk,leclair,bernard} 
Random matrix ensembles with this symmetry,
representing the zero-dimensional limit appropriate for quantum dots,
have a density of states that vanishes quadratically in energy
at energies much smaller than the mean level spacing. \cite{alt-zirn,zirn-su2}
Similar behaviour is expected for
finite-dimensional systems if states are Anderson localised,
on the grounds that random matrix theory should describe
states within a localisation volume. \cite{sent-fish}
Calculations for one-dimensional systems from class C,
using either supersymmetry \cite{sigma1,sent-fish} 
or the DMPK equation, \cite{dmpk} confirm this idea.

Existing information on localisation in class C systems is also provided 
in part by
renormalisation group treatments \cite{sigma} of the appropriate 
non-linear sigma model at weak coupling, corresponding to 
weak disorder. These calculations identify two as the lower critical
dimension. Thus, as in the Wigner-Dyson symmetry classes,
it is only in more than two dimensions that
a transition occurs between localised and metallic phases as a
function of disorder strength, while in one dimension
even weak disorder is sufficient to localise all states.
In addition, 
for two-dimensional systems
with broken time-reversal symmetry, including both ones
from the Wigner-Dyson unitary class and ones from class C,
a delocalisation transition of the quantum Hall type is possible,
and it is this transition that has been the focus of past
work on class C network models. \cite{kag-prb,kag-prl,grl,smf,cardy-sqhe}

Many of these aspects, including the form of 
the density of states in a localised 
phase and the possibility of a
quantum Hall plateau transition, emerge naturally from the 
approach we describe here, which is presented as follows.
In Sec.\,\ref{models} we introduce in detail the models 
that we are concerned with. In Sec.\,\ref{results} 
we set out our general results, relating the density of
states and average conductance for a network model to 
averages over certain classical random walks, 
and present proofs of these results.
We describe applications of these general results
to the spin quantum Hall effect, to random matrix theory,
and to localisation on a Cayley tree, in Sec.\,\ref{applications}.
Open questions and future prospects are discussed in Sec.\,\ref{discussion}.

\section{Models}
\label{models}

We shall be concerned with models both for closed systems
and (in connection with the Landauer formula for conductance)
for open systems, but we restrict definitions initially to
closed systems.
Consider a graph $\cal G$ consisting of directed edges $e$ connecting nodes
$n$, each of degree four, with the restriction that 
at each node two directed edges enter and two leave.
An $N$-component wavefunction propagates on each edge.
This propagation may be described by a unitary
evolution operator $\cal U$, which evolves the wave function one unit
forwards in time, as the particle moves from a given edge to a
neighbouring one. The evolution operator plays the same role in defining
the network model as does the Hamiltonian in the case, for example, of
a tight-binding model. It has been discussed 
for the $U(1)$ network model in Refs.\,\onlinecite{kless-metz} and 
\onlinecite{ho-chalk}. 
In general, it is constructed from two ingredients. First, with
each edge $e$ is associated a unitary $N\times N$ matrix $U_e$.
This matrix specifies the (generalised) phase acquired on 
traversing the link. Second, with each node $n$ is associated an
$S$-matrix of the form 
\begin{equation} 
S_n=
\openone \otimes \left(\begin{array}{cc}\cos\theta_n & \sin\theta_n \\
                      -\sin\theta_n & \cos\theta_n \end{array}\right)\,,
\end{equation}
where $\openone$ is the $N\times N$ unit matrix.
This $S$-matrix describes scattering at the node from the 
incoming edges to the outgoing ones.
If $\cal G$ has $E$ edges (and therefore $E/2$ nodes), then $\cal U$
is an ${\cal N}\times {\cal N}$ matrix, with ${\cal N}=EN$.
It consists of $E/2$ blocks, each associated with a particular node
and of size $2N \times 2N$. The block at the node $n$ has the form
\begin{equation}
\left(\begin{array}{cc}U_3^{1/2}&0\\0&U_4^{1/2} \end{array}\right)\,
S_n
\,\left(\begin{array}{cc}U_1^{1/2}&0\\0&U_2^{1/2} \end{array}\right)
\label{node}
\end{equation}
where $(1,2)$ and $(3,4)$ label the edges which are 
respectively incoming and outgoing at this node.

So far, the symmetry class of the network model has not been fixed,
except that propagation along directed links breaks time-reversal symmetry.
To identify network models from class C, one starts\cite{kag-prl}
from the defining property of a Hamiltonian $\cal H$ with this symmetry,
which is \cite{alt-zirn}
\begin{equation}
\label{class-C}
{\cal H}^* = - \sigma_y {\cal H} \sigma_y\,,
\end{equation}
where $\sigma_y$ denotes the conventional Pauli matrix acting on
spin variables and ${\cal H}^*$ is the complex conjugate
of $\cal H$. Applying this to $\cal U$, interpreted as 
${\cal U} = e^{i{\cal H}}$,
the number of wavefunction components $N$ must be even, so that
the space of states on each link may be viewed as consisting of
$N/2$ two-component subspaces, within which $\sigma_y$ operates. Then Eq.\,(\ref{class-C}) becomes
\begin{equation}
\label{class-C'}
{\cal U}=\sigma_y\, {\cal U}^*\sigma_y\,,
\end{equation}
and from this an equivalent restriction follows on the edge
phases, ${U_e}=\sigma_y\, {U_e}^*\sigma_y$, which are therefore 
unitary Sp($N$) matrices,  equivalent for $N=2$ to SU(2) matrices.

Randomness is introduced into these models via the edge phases.
We take them to be independent random variables 
drawn from a distribution which is uniform on the invariant
(Haar) measure of Sp($N$).
The quenched average of a given quantity in the network model, 
denoted by $\langle\ldots\rangle$, 
is the mean with respect to this measure.

An open system is constructed from a closed system of this type
by `cutting open' a number of edges. The two halves of each edge 
cut into two in this
way constitute one new edge directed into the system 
and one new edge directed out of the system.
We may consider a conductance experiment between two `contacts' by grouping
a subset of these of the incoming edges $\{e_{\rm in}\}$
to form one contact and
another subset of the outgoing edges $\{e_{\rm out}\}$ to form the other.
The transmission matrix $t$ between these two contacts is a rectangular
matrix whose elements are $\langle e_{\rm out}|(1-{\cal U})^{-1}|e_{\rm
in}\rangle$.
The spin conductance measured between the
two contacts in units of $(\hbar/2)^2/h$ is
\begin{equation}
\label{conductance}
g = {\rm Tr}\, t^{\dagger}t\,.
\end{equation}
from the multi-channel Landauer formula.

Clearly, a great variety of specific models can be constructed within
this framework, by making different choices for the graph $\cal G$ and for
the number of channels $N$. We defer discussion of particular examples
to Sec.\,\ref{applications}. 

\section{General results}
\label{results}

In this section we state and prove our results for a general graph $\cal G$.
We consider first the particular case of $N=2$, and discuss the extension to
general $N$ in Sec.\,\ref{general-N}.

\subsection{Green function, Feynman path expansion and classical walks}
\label{walks}

The Green function for propagation from edge $e'$ to edge $e$
is (with $N=2$) a $2\times 2$ matrix
\begin{equation}
\label{Gee}
G(e,e';z)\equiv \langle e|(1-z{\cal U})^{-1}|e'\rangle
\end{equation}
where $|e\rangle$ is a state in the two-component space of wavefunctions for 
a particle located on the edge $e$. For $|z|<1$ Eq.\,(\ref{Gee}) 
may be expanded as a sum over Feynman
paths on $\cal G$ which begin on $e'$ and end on $e$: each path
gives an ordered product of factors $zU_j$ along edges traversed by the 
path, weighted by appropriate factors 
of $\cos\theta_n$ and $\pm\sin\theta_n$ for each node through which it
passes. Alternatively, we may re-write Eq.\,(\ref{Gee}) as
\begin{equation}
G(e,e';z)= -\langle e|z^{-1}{\cal U}^{\dagger}
(1-z^{-1}{\cal U}^{\dagger})^{-1}|e'\rangle\,,
\end{equation}
obtaining instead a series convergent for $|z|>1$,
involving an ordered product of factors $z^{-1}U^{\dagger}_j$
and an overall negative sign for each Feynman path.

\begin{figure}
\includegraphics[width=8cm]{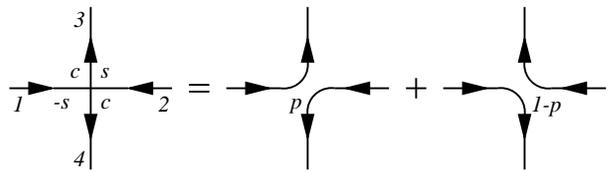}
\caption{Decomposition of a given node. In the network model, 
$S$-matrix elements
$\cos\theta_n$ and $\pm\sin\theta_n$ are associated with the transitions
$(1,2)\to(3,4)$ as indicated on the left. Each decomposition of the 
node is then
weighted with factors $p_n=\cos^2\theta_n$ and $1-p_n=\sin^2\theta_n$ as
indicated.}
\label{decomp}
\end{figure}

Our central result is an expression for the disorder-average, 
Sp(2) trace of the Green function, ${\rm Tr}\,\langle G(e,e,z) \rangle$, 
in terms of classical paths. To state this result, we define
on the same graph $\cal G$ a \em classical \em scattering
problem as follows. Each node may be decomposed into two disconnected
pieces in two ways, viz.~$(13,24)$ or $(14,23)$, as illustrated in Fig.~\ref{decomp}. 

\noindent{\bf Theorem 1.} 
The average Green function
is given for $|z|<1$ by the generating function for the probability $P(e;L)$
in the classical problem that the edge $e$
belongs to a loop of a given length $L$. Explicitly
\begin{equation}
\label{Gee-avge1}
{\rm Tr}\,\langle G(e,e;z)\rangle = 2-\sum_{L>0}P(e;L)z^{2L}\,.
\end{equation}
For $|z|>1$ it is given instead by
\begin{equation}
\label{Gee-avge2}
{\rm Tr}\,\langle G(e,e;z)\rangle = \sum_{L>0}P(e;L)z^{-2L}\,.
\end{equation}

\subsection{Density of states}

The eigenvalues of the evolution operator $\cal U$ for a closed graph lie on
the unit circle in the complex plane and may be written as $\exp(i\epsilon_j)$,
with eigenphases $-\pi < \epsilon_j \leq \pi$ for $j=1 \ldots {\cal N}$. These
eigenphases are analogous for the network model to the energy eigenvalues of
a system specified by its Hamiltonian. We define the density of states to be
\begin{equation}
\rho(\epsilon)\equiv \frac{1}{\cal N} \sum_j 
\langle \delta(\epsilon - \epsilon_j) \rangle\,.
\end{equation}
A consequence of the symmetry of Eq.\,(\ref{class-C'}) is that 
$\rho(\epsilon)=\rho(-\epsilon)$.

Defining $P(L)$, the edge-average of $P(e;L)$, by
\begin{equation}
P(L)=\frac{1}{E}\sum_e P(e;L)\,,
\end{equation}
it follows from Eqns.\,(\ref{Gee-avge1}) and (\ref{Gee-avge2}) that 
\begin{equation}
\label{rho}
\rho(\epsilon) = 
\frac{1}{2\pi}\left[1 - \sum_{L>0}P(L)\cos(2L\epsilon)\right]\,.
\end{equation} 

\subsection{Conductance}

The classical scattering problem introduced in Sec.\ref{walks} may also be 
considered for an open system. For an open system with $M$ cut links,
each decomposition breaks $\cal G$ into $M$ directed paths
which each run from an ingoing edge to an outgoing one, together with
a number (possibly zero) of closed loops. Let $P(e,e')$ be the probability 
that a path runs from the ingoing edge $e'$ to the outgoing edge $e$.

\noindent{\bf Theorem 2.} 
The disorder-average of the conductance defined in Eq.\,(\ref{conductance})
is given by
\begin{equation}
\langle g \rangle = 2\sum_{e \in 1, e' \in 2} P(e,e')\,,
\end{equation}
where the sets $1$ and $2$ denote respectively: the edges incident on 
the first contact from $\cal G$; and those incident on $\cal G$ from the 
second contact.

\subsection{Proofs}

It is useful to introduce for a closed system the resolvent 
\begin{equation}
R(z)\equiv\sum_e{\rm Tr}\,G(e,e;z)
\end{equation}
(where Tr again indicates an Sp(2) trace)
and to generalise this to the case where the parameter
$z$ takes independent values $z_e$ on each edge $e$. 
The expansion of $R(\{z\})$
as a sum over paths then yields a multinomial expression in all the $z_e$.

We now require two Lemmas about Sp$(2)$
matrices.

\noindent{\bf Lemma 1.} If $U\in{\rm Sp}(2)$, then its mean $q$th
moment, $\langle U^q\rangle$, is zero unless $q=0$ or $q=2$, in which
case it takes the value $\bf 1$ or $-\frac12\bf 1$ respectively. This can
be shown using the representation 
$U=\exp(i\alpha\vec n\cdot\vec{\bf\sigma})$, where
$\vec n$ is a unit real 3-vector and the $\bf\sigma$s are the Pauli
matrices, so that
\begin{equation} 
U^q= \cos q\alpha\,{\bf 1}+i\sin q\alpha\,\vec n\cdot\vec{\bf\sigma}\,.
\end{equation}
The result now follows from the observation that the invariant measure for
Sp$(2)$ is the uniform measure on the group manifold S$_3$, and
therefore has the form $\int dU=\pi^{-1}\int_0^\pi(1-\cos2\alpha)d\alpha\int
d\vec n$.

\noindent{\bf Lemma 2.} If $G$ is a real linear combination of Sp$(2)$
matrices, it is itself proportional to an Sp$(2)$ matrix, with a real 
scalar constant of proportionality.
This follows directly from the above representation for each matrix.

The main argument in the proof of theorem 1 now proceeds as follows. 
As discussed above, for each
realisation of the randomness, $R(\{z\})$ is
a sum over closed directed paths on $\cal G$. 
For a particular path,
each link or node may be traversed an arbitrary
number of times. We first show that it is sufficient, in calculating
the mean $\langle R(\{z\})\rangle$, to restrict this sum to those paths
which traverse each edge exactly twice or not at all. 
Let us consider the sum of all paths in which
a particular edge $e$ is traversed exactly $q$ times. This has the form
\begin{equation}
\label{UAq}
z_e^q{\rm Tr}\,[U_eA(e,e)]^q
\end{equation}
where $A(e,e)$ denotes the sum 
over \em all \em weighted paths which begin and end
on $e$, but do not themselves traverse $e$.
By Lemma 2, this is proportional to an Sp$(2)$ matrix, \cite{footnote}
so it may be written as 
\begin{equation}
A(e,e)=|A(e,e)|\widetilde A(e,e)
\end{equation}
where $|A(e,e)|$ is real and $\widetilde A(e,e)\in{\rm Sp}(2)$.
Defining $U_e'\equiv U_e\widetilde A(e,e)$, (\ref{UAq}) may
e written
\begin{equation}
z_e^q{\rm Tr}\,{U_e'}^q|A(e,e)|^q\,.
\end{equation}
The invariant integration over $U_{e}$ is equivalent to that over
$U_e'$, so that, by Lemma 1, the result will vanish unless $q=0$ or
$q=2$. Since $\langle R(\{z\})\rangle$ is a multinomial 
expression in the 
parameters $\{z\}$, the argument may be applied to each edge in
turn to show that the only allowed powers of any $z_e$ entering this expression
are 0 or 2.
This establishes the first part of the proof.

For paths in which 
each node is visited only 0 or 2 times, the main result follows
immediately. For such a path must 
traverse a closed loop in the decomposition of $\cal G$ exactly
twice. At each node there will be factors of $\cos^2\theta_n$ or
$(\pm\sin\theta_n)^2$, giving precisely the correct weighting for this
loop to appear in the decomposition. The product of Sp$(2)$ matrices
along the loop will have the form
\begin{equation}
{\rm Tr}\, \langle(U_1U_2\ldots)(U_1U_2\ldots)\rangle
\end{equation}
Thus, defining $U_1'\equiv U_1U_2\ldots$, this is equivalent to
averaging ${U_1'}^2$, which gives $-\frac12$ by Lemma 1. Finally, the
trace gives a factor of 2.

Further effort is required to treat paths which visit some 
nodes more than twice. A little thought shows that such a 
node must be visited exactly
four times, entering and leaving exactly twice along each directed edge,
if the contribution to $R(z)$ is not to vanish on averaging.
Consider such a node $n$, and label the incoming and outgoing edges as $(1,2)$,
$(3,4)$ respectively (see Fig.~\ref{decomp}). 
We show that the sum over all paths
visiting this node four times may be written in terms of the two ways of
decomposing this node, with precisely the correct weights. 
Let $A(i,j)$ be the sum
over all paths from edge $j\in(3,4)$ to edge $i\in(1,2)$, 
which do not use any of these four edges. The sum over paths visiting the 
node four
times may be decomposed into
eighteen different contributions, depending on the order
in which the edges are visited. Each contains a product of
four factors $A(i,j)$ as
well as Sp$(2)$ matrices $U_i$ and $U_j$.
There are in fact two types of contribution: in the first type,
two of the $A(i,j)$
appear twice and the others not at all;
while in the second type all four $A(i,j)$
appear once each. There are six of
the first type and they have the form
\begin{widetext}
\begin{eqnarray}
&&{\rm Tr}\,U_1A(1,3)U_3U_2A(2,4)U_4U_1A(1,3)U_3U_2A(2,4)U_4\,s^4\label{mes1}\\
&&{\rm Tr}\,U_1A(1,3)U_3U_1A(1,3)U_3U_2A(2,4)U_4U_2A(2,4)U_4\,c^2(-s^2)\label{mes2}\\
&&{\rm Tr}\,U_1A(1,3)U_3U_2A(2,4)U_4U_2A(2,4)U_4U_1A(1,3)U_3\,c^2(-s^2)\label{mes3}
\end{eqnarray}
and
\begin{eqnarray}
&&{\rm Tr}\,U_1A(1,4)U_4U_2A(2,3)U_3U_1A(1,4)U_4U_2A(2,3)U_3\,c^4\label{mes4}\\
&&{\rm Tr}\,U_1A(1,4)U_4U_1A(1,4)U_4U_2A(2,3)U_3U_2A(2,3)U_3\,c^2(-s^2)\label{mes5}\\
&&{\rm Tr}\,U_1A(1,4)U_4U_2A(2,3)U_3U_2A(2,3)U_3U_1A(1,4)U_4\,c^2(-s^2)\label{mes6}
\end{eqnarray}
\end{widetext}
where we have introduced the shorthand $s\equiv\sin\theta_n$ and
$c\equiv\cos\theta_n$. (Note that (\ref{mes2}) and (\ref{mes3}) give
separate contributions to $R(\{z\})$;
similarly for (\ref{mes5}) and (\ref{mes6})).
As before, write $A(i,j)=|A(i,j)|\widetilde A(i,j)$ and note that, by
a change of integration variable, (\ref{mes1}) is, on averaging,
equivalent to the average of $U_1^2$ (that is, $-\frac12$), multiplied
by $|A(1,3)|^2|A(2,4)|^2s^4$. By a similar argument, each of
(\ref{mes2}) and (\ref{mes3}) are equal to the average of $U_1^2U_2^2$
(that is, $(-\frac12)^2$), multiplied by $|A(1,3)|^2|A(2,4)|^2c^2(-s^2)$.
The total of the (\ref{mes1}) - (\ref{mes3}) is therefore
\begin{equation}
\begin{split}
\label{first-one}
&
-|A(1,3)|^2|A(2,4)|^2\left(s^4+2(-\tfrac{1}{2})c^2(-s^2)\right)\\
&
=-|A(1,3)|^2|A(2,4)|^2\,\sin^2\theta_n
\end{split}
\end{equation}
Similarly, (\ref{mes4}-\ref{mes6}) sum up to
\begin{equation}\label{first-two}
-|A(1,4)|^2|A(2,3)|^2\,\cos^2\theta_n
\end{equation}

Now consider the other twelve contributions, in which $A(1,3)$, $A(2,4)$,
$A(1,4)$ and $A(2,3)$ each appear exactly once. Six of these are
\begin{widetext}
\begin{eqnarray}
&&{\rm Tr}\,U_1A(1,3)U_3U_1A(1,4)U_4U_2A(2,4)U_4U_2A(2,3)U_3\,c^4\label{mess1}\\
&&{\rm Tr}\,U_1A(1,3)U_3U_2A(2,3)U_3U_2A(2,4)U_4U_1A(1,4)U_4\,s^4\\
&&{\rm Tr}\,U_1A(1,3)U_3U_1A(1,4)U_4U_2A(2,3)U_3U_2A(2,4)U_4\,c^2(-s^2)\\
&&{\rm Tr}\,U_1A(1,3)U_3U_2A(2,3)U_3U_1A(1,4)U_4U_2A(2,4)U_4\,c^2(-s^2)\\
&&{\rm Tr}\,U_1A(1,3)U_3U_2A(2,4)U_4U_1A(1,4)U_4U_2A(2,3)U_3\,c^2(-s^2)\\
&&{\rm Tr}\,U_1A(1,3)U_3U_2A(2,4)U_4U_2A(2,3)U_3U_1A(1,4)U_4\,c^2(-s^2)\label{mess2}
\end{eqnarray}
\end{widetext}
In addition, there are six equal contributions in which the factors are
cyclically permuted so that each begins $U_1A(1,4)\ldots$. 
Each term is proportional to $|A(1,3)||A(2,4)||A(1,4)||A(2,3)|$.
The remainder of the expressions may be simplified, for example, by redefining
$U_1\widetilde A(1,3)=U_1'$ and 
$U_2\widetilde A(2,4)=U_2'$. This reduces (\ref{mess1}-\ref{mess2}) to
\begin{eqnarray}
&&{\rm Tr}\,U_1'U_3U_1'BU_4U_2'U_4U_2'CU_3\,c^4\label{a}\\
&&{\rm Tr}\,U_1'U_3U_2'CU_3U_2'U_4U_1'BU_4\,s^4\label{b}\\
&&{\rm Tr}\,U_1'U_3U_1'BU_4U_2'CU_3U_2'U_4\,c^2(-s^2)\label{c}\\
&&{\rm Tr}\,U_1'U_3U_2'CU_3U_1'BU_4U_2'U_4\,c^2(-s^2)\label{d}\\
&&{\rm Tr}\,U_1'U_3U_2'U_4U_1'BU_4U_2'CU_3\,c^2(-s^2)\label{e}\\
&&{\rm Tr}\,U_1'U_3U_2'U_4U_2'CU_3U_1'BU_4\,c^2(-s^2)\label{f}
\end{eqnarray}
where $B\equiv\widetilde A(1,3)^{-1}\widetilde A(1,4)$
and $C\equiv\widetilde A(2,4)^{-1}\widetilde A(2,3)$.
By making suitable changes of integration variables, as before, we
find that
(\ref{a}) and (\ref{b}) give factors of $(-\frac12)^2
{\rm Tr}\,(BC)$, while 
(\ref{c}) - (\ref{f}) give $(-\frac12)^3{\rm Tr}\,(BC)$.
The sum of all twelve such contributions is therefore
\begin{equation}\begin{split}
X&\equiv\dfrac{1}{2}{\rm Tr}\,(BC)|A(1,3)||A(2,4)||A(1,4)||A(2,3)|\\
&\times\,(c^4+s^4+2c^2s^2)\\
&=\dfrac{1}{2}{\rm Tr}\,A(1,3)^{\dag}A(1,4)A(2,4)^{\dag}A(2,3)
\end{split}
\end{equation}
The important feature of this result is that it is \em independent
\em of $\theta_n$. 
It may be written, trivially, as
\begin{equation}
X= p_nX + (1-p_n)X
\end{equation}
while the first six contributions have the form 
(as given in \ref{first-one} and \ref{first-two})
\begin{equation}
-p_n|A(1,4)|^2|A(2,3)|^2-(1-p_n)|A(1,3)|^2|A(2,4)|^2
\end{equation}
Therefore we can obtain the same total result, after averaging over
$U_1,\ldots,U_4$, if we decompose the node $n$
in each of the two possible ways, and weight the
two decompositions with probabilities
$p_n$ and $1-p_n$ respectively (see Fig.~\ref{decomp}). 
For in the first case, $(13,24)$,
we find $-|A(1,4)|^2|A(2,3)|^2+X$, while in the second case of
$(14,23)$, we get $-|A(1,3)|^2|A(2,4)|^2+X$. 

Now we may simply go through $\cal G$, decomposing it node by node. 
The result is that $\langle R(\{z\})\rangle$ is given by the
weighted sum over the same quantity calculated on graphs arising from 
all possible decompositions of $\cal G$; such graphs consist of
closed loops with no remaining nodes. In
each decomposition, a given edge $e$ 
belongs to just one loop, and, for that loop, of length $L$ say,
the contribution to ${\rm Tr}\,G(e,e;z)$ is
just $-z^{2L}$. Thus the mean 
${\rm Tr}\,\langle G(e,e;z)$ is given by the average of
this quantity over the ensemble of loops corresponding to the decomposition of
$\cal G$. This completes the proof of Theorem 1.

So far we have considered unitary evolution on a closed system. However, since the
proof is constructed to work for arbitrary values of the 
parameters $\{z_e\}$, it
generalises straightforwardly to an open system where probability is not 
conserved at the nodes which are connected to external leads. 
For example, if in
Fig.~\ref{decomp}, the edges $(1,3)$ correspond to external leads which
carry no current, then we should
consider only those Feynman paths through $n$ which pass directly 
from $4\to2$, with amplitude
$\cos\theta_n$. In the above argument, 
this may be taken into account by regarding
this node as a single edge, carrying an Sp$(2)$ matrix $U_4U_2$, and with fugacity
$z_2z_4\cos\theta_n$. With this modification, the proof of Theorem
1 goes through as before.

Now consider conductance measurements on a open graph ${\cal G}$, with
external contacts $\{e_{\rm in}\}$ and $\{e_{\rm out}\}$, as described
in Sec.~\ref{models}. Consider the graph $\cal G'$ formed from 
${\cal G}$ by joining particular outgoing and incoming external edges
$e_{\rm out}$ and $e_{\rm in}$ to create a new internal edge $e$.
Observe that, in
calculating $\langle G(e,e;z)\rangle$, the edge $e$ is always traversed exactly twice. 
Therefore 
\begin{equation}
\label{GG2}
\langle G(e_{\rm out},e_{\rm in};z)^2\rangle_{\cal G}=
\langle G(e,e;z)\rangle_{\cal G'}\Big|_{z_e=1}
\end{equation}
Since $G(e_{\rm out},e_{\rm in})$ is given by a sum over Sp$(2)$
matrices,
by Lemmas 1 and 2, $\langle G(e_{\rm out},e_{\rm in};z)^2\rangle=-\frac12
\langle|G(e_{\rm out},e_{\rm in})|^2\rangle$.
Now apply Theorem 1: it follows that 
$\langle|G(e_{\rm out},e_{\rm in})|^2\rangle$ is twice the probability
that $e_{\rm in}$ and $e_{\rm out}$ are connected by a path in the
decomposition of $\cal G$. Summing over all the edges $e_{\rm in}$ and
$e_{\rm out}$ as required by the Landauer formula (\ref{conductance})
then gives the result stated in Theorem 2.

\subsection{General $N$}
\label{general-N}

Our approach to the Sp($N$) model with even $N>2$ is based on the fact that a general Sp($N$) rotation can be written as a product of 
rotations derived from a limited number of generators. Moreover, under rather weak conditions,
the probability distribution of such a product containing random
factors will converge as the number of factors increases
to the Haar measure on Sp($N$): the requirement
is simply that no subspace is left invariant by the ensemble of rotations.

\begin{figure}
\includegraphics[width=8cm]{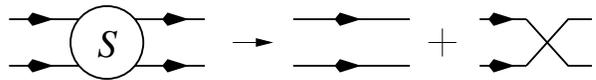}
\caption{Implementation of a link in the Sp(4) model. It consists of
pairs of incoming and outgoing Sp(2) links, 
connected through
a random $S$-matrix which mixes the two channels. After
averaging, this is equivalent to the decomposition shown on the right
hand side, equally weighted.}
\label{Sp4}
\end{figure}

We therefore 
build the Sp($N$) model on $\cal G$ by taking $N/2$ copies of the
Sp(2) model, each individually defined on $\cal G$, 
and coupling the copies together.
This coupling takes the form of a product of many non-commuting
matrices. Successive factors in the product are of two kinds.
One kind is block-diagonal with $N/2$ random Sp(2) blocks,
resulting in intra-copy scattering; 
the other kind must produce inter-copy scattering and may
be chosen non-random. 
It is convenient to restrict the inter-copy scattering to be between 
corresponding links in each copy. 
For example, in the case of Sp(4) each pair of corresponding links 
is coupled as shown in Fig.~\ref{Sp4}. The evolution matrix for
this pair of links consists of a product $S$ of a large number of factors, each of the form
\begin{equation}
\frac{1}{\sqrt 2}\left(\begin{array}{cc}U_1&0\\0&U_2 \end{array}\right)\,\cdot
\left(\begin{array}{cc}{\bf 1} & {\bf 1} \\
                     -{\bf 1} & {\bf 1}\end{array}\right)\,,
\end{equation}
where $U_1$ and $U_2$ are random Sp(2) matrices, chosen independently
for each term in the matrix product.

The advantage of this choice is that we may immediately apply our main
theorem to show that, after averaging over the random Sp(2) matrices,
the mean density of states and mean conductance are given in terms of
a decomposition of each node as before, and a decomposition of each
link as shown in Fig.\ref{Sp4}, with equal probabilities for each term.
This construction may be generalised to Sp($N$) for arbitrary even $N$:
the result is that, in the classical system, the $N/2$ exit trajectories
are a random permutation of the incident ones, with all permutations
having equal weight, independently for each edge.

\section{Applications of general results}
\label{applications}

There are three obvious classes of behaviour possible in the classical
scattering problem which we have arrived at, corresponding to loops,
or closed classical walks, that are localised, extended or critical.

By {\em localised} classical walks, we mean that only a vanishing fraction
of walks have infinite length, and that the number of walks longer than a 
characteristic size $\xi$ decreases rapidly with $\xi$. One might
have, for example, $P(L) \sim \exp(-L/\xi)$ for $L\gg \xi$.
Then the fraction of loops longer than $L$, given by
\begin{equation}
f(L)= 1-\sum_{l=1}^{L} P(l)\,,
\end{equation}
approaches zero as $L \to \infty$. As a consequence, 
$\rho(\epsilon) \to 0$ as $\epsilon \to 0$. 
Introducing the mean square length
of loops,
\begin{equation}
\langle L^2 \rangle = \sum_{L>0} P(L) L^2\,,
\end{equation}
then provided $\langle L^2 \rangle$ is finite, $\rho(\epsilon)$
vanishes quadratically, varying as
\begin{equation}
\rho(\epsilon) = (\pi)^{-1} \langle L^2 \rangle \epsilon^2\,+{\cal O}(\epsilon^4)\,.
\end{equation}
This is the behaviour expected in the localised phase of
the quantum problem. Moreover, if classical walks have a characteristic
size $\xi$ and if $\cal G$ is embedded in
Euclidean space, then we expect the conductance to decrease rapidly
with increasing contact separation, for separations larger than $\xi$.

Alternatively, classical walks are {\it extended} 
if a finite fraction have infinite length, so that 
$\lim_{L \to \infty} f(L) > 0$. In this case, $\rho(0)=(2\pi)^{-1}f(\infty)$
is non-zero, which we expect in the extended phase of the quantum problem.
Clearly, the form of $\rho(\epsilon)$ at small $\epsilon$ is determined
by that of $f(L)$ at large $L$: if $f(L)-f(\infty) \sim L^{-x}$
with $0<x<2$ then $\rho(\epsilon)-\rho(0) \propto |\epsilon|^x$,
while if $f(L)-f(\infty)$ falls faster than $L^{-2}$ at large $L$,
then $\rho(\epsilon)-\rho(0) \sim \epsilon^2$. 
It is an interesting consequence of Eq.~(\ref{rho}) that $\rho(\epsilon)$
is bounded, and so a divergence in the density of states, as occurs
for example in class D localisation problems, is excluded in class C
network models, whatever the choice for $\cal G$.
In the extended phase for a system in $d$-dimensional
Euclidean space, Ohm's law dictates that
the two-terminal conductance should vary with the separation, $l_{\parallel}$,
between terminals and their cross-sectional area, $l_{\perp}$, 
as $l_{\perp}^{d-1}/l_{\parallel}$ for large $l_{\parallel}$, $l_{\perp}$;
this places strong constraints on behaviour in the 
corresponding ensemble of classical walks, suggesting that, on large
distance scales, they should behave like free random walks. 

Classical walks that are {\it critical} have no characteristic loop
size, and a vanishing fraction that are of infinite length.
Then $f(\infty)=0$ and a possible behaviour for $P(L)$ is
$P(L)\sim L^{-y}$ at large $L$, with $y>1$. 
The resulting quantum density
of states has the critical behaviour
$\rho(\epsilon)\sim |\epsilon|^{y-1}$ for small $\epsilon$.
The conductance of such a system is expected to depend on the geometry of
sample and contacts, but to be unchanged under a rescaling of
all spatial dimensions.

The task that remains, given a particular model for quantum localisation,
specified by the graph $\cal G$ and values for the node probabilities $p_n$,
is to determine behaviour of the corresponding classical walks. In general, 
this remains a challenging open problem with connections
to previously-studied random-walk problems which we summarise in 
Sec.~\ref{discussion}. In particular instances, however, relevant properties 
of the classical walks can be calculated; we describe three examples in
the remainder of this section.

\subsection{Spin quantum Hall effect}

\begin{figure}
\includegraphics[width=7cm]{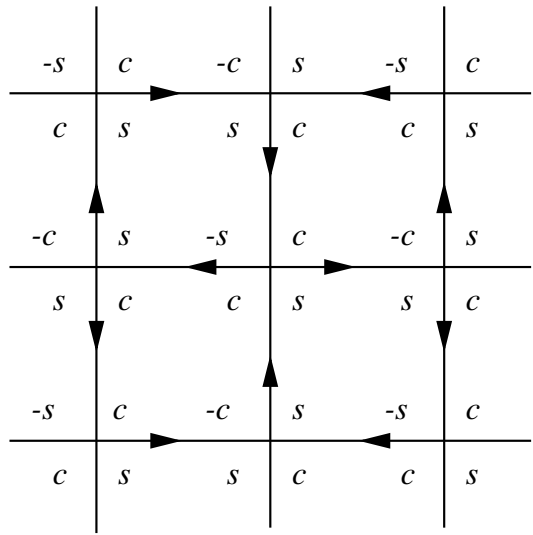}
\caption{The network model for the spin quantum Hall effect,
defined using the L-lattice. Scattering amplitudes of
$\pm\cos\theta_n$ and $\pm\sin\theta_n$ are indicated
by $\pm c$ and $\pm s$.}
\label{L-lattice}
\end{figure}
A two-dimensional model exhibiting the spin quantum Hall effect is obtained
by taking $\cal G$ to be the L-lattice, illustrated in Fig.\,\ref{L-lattice}.
As GRL have shown, the two possible classical decompositions
of a node may be associated with the presence or absence of a
bond, with probabilities $p$ and $1-p$,
between neighbouring sites on an associated square lattice.
This associated lattice is rotated by $45^{\circ}$ relative to the
L-lattice, and has a larger lattice spacing by a factor $\sqrt 2$.
In this way, closed loops of the classical problem form
interior or exterior
hulls of bond percolation clusters on the larger lattice.
It is known that such loops are finite with characteristic size
$\xi$ except at the critical point, $p=p_c$, which for bond percolation on the
square lattice occurs at $p_c=1/2$. On approaching the critical point,
$\xi$ diverges as $\xi \sim |p-p_c|^{-\nu}$ with $\nu=4/3$, 
while at the critical point
the distribution of hull lengths is $P(L)\sim L^{-y}$ at large $L$, with $y=8/7$.
In this way one finds that the quantum localisation length diverges
with the same exponent value, $\nu=4/3$, as the plateau transition
is approached, and that the density of states $\rho(\epsilon)$
varies for small $\epsilon$ as $\rho(\epsilon) \sim \epsilon^2$
in the localised phase, and as $\rho(\epsilon) \sim |\epsilon|^{1/7}$
at the critical point.

\subsection{Random matrix theory}

The simplest application of our discussion of Sp($N$) models
with general $N$ is to random matrix theory. To this end, we take $\cal G$
to consist of a single edge closed on itself. Then the evolution
operator $\cal U$ is a random Sp($N$) matrix, chosen with
the Haar measure. In this case the density of states $\rho(\epsilon)$
is the eigenphase density for the Sp($N$) random matrix ensemble,
which has been determined previously by Zirnbauer\cite{zirn-su2}
using supersymmetry methods. 

Applying to this problem the approach we have described in 
Sec.~\ref{general-N}, 
we must consider $N/2$ copies of an edge, which are
closed by connecting outgoing ends to a permutation of ingoing ends. 
For example, the case $N=4$ corresponds to joining opposite ends of the
two possibilities shown in Fig.~\ref{Sp4}, thus producing with equal
probability either two loops, each of length 1, or a single loop of length 2.
For general $N$, all possible lengths $L$ up to $N/2$ are possible, with
equal probability. Thus
$P(L)=2/N$ for $1\leq L \leq N/2$,
and $P(L)=0$ otherwise. Using (\ref{rho}) we thence find
\begin{equation}
\rho(\epsilon)=\frac{N+1}{2\pi N}
\left[1-\frac{\sin(N+1)\epsilon}{(N+1)\sin\epsilon}\right]
\end{equation}
in agreement with Ref.~\onlinecite{zirn-su2}.

\subsection{Cayley tree}

\begin{figure}
\includegraphics[width=7cm]{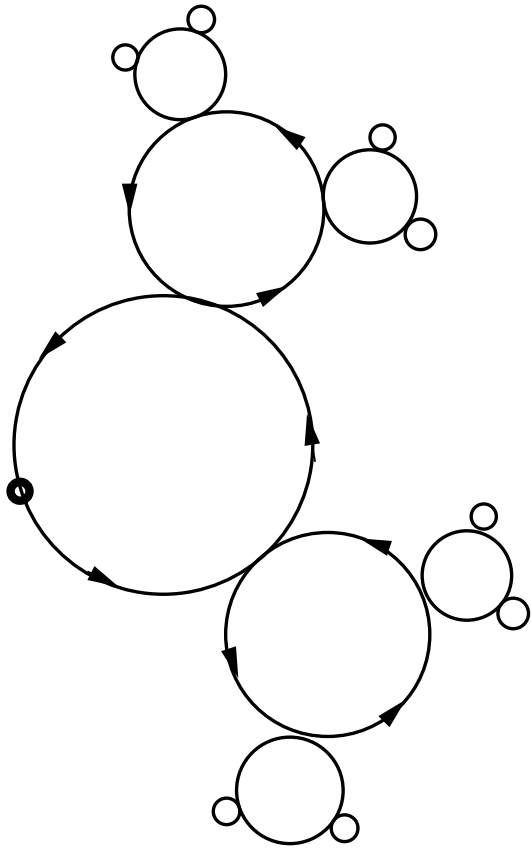}
\caption{The graph $\cal G$ for network model based on
a rooted Cayley tree. An example with coordination number $q=3$
and four generations is shown; system size is 
increased by increasing the number of generations.
S-matrix amplitudes at nodes are $\cos\theta$
to remain at the same generation, and $\pm \sin\theta$ 
to change generations.}
\label{cayley}
\end{figure}
A solvable model which illustrates each of the types of behaviour - localised, extended and critical - for classical loops is based on 
the 
geometry of the Cayley tree. Specifically, we take $\mathcal{G}$ to be a graph of the type illustrated in Fig 4; the 
$U(1)$ network model on such a tree has been studied previously in Ref.~\onlinecite{chalk-siak}. We restrict attention to $N=2$ and 
consider first 
coordination number $q=3$. As previously, we define $P(e;L)$ to be the probability that the edge $e$ of the root, far from the surface 
of the 
tree, lies on a given closed loop of length $L$.\\
Then
\begin{equation} \label{eb} \begin{split}
P(e;3L)&=p^2\delta_{L,1}+2p(1-p) P(e;3L-3)\\ + 
&(1-p)^2 \sum_{m=1}^{L-2} P(e;3m)P(e;3L-3m-3)\,.
\end{split}\end{equation} 
This can be solved by using the generating function $G(z)=\sum_{r\geq1}z^rP(e;3r)$: we obtain 
\begin{equation}G(z)=zp^2+2zp(1-p)G(z)+(1-p)^2zG(z)^2\,.\end{equation}
\\We take the branch of $G(z)$ such that $G(0)=0$; thus 
\begin{equation}G(z)=\frac{1}{2z(1-p)^2}\left(1-2zp(1-p)-\sqrt{1-4zp(1-p)}\right)\,.\end{equation}
When $z=1$, $\sqrt{1-4zp(1-p)}$ is equal to $1-2p$ for $p<1/2$ and $2p-1$ for $p>1/2$. Therefore for $p>1/2$, $G(1)=1$, so 
that all walks are localised. The characteristic loop size is given by
\begin{equation}<L>\, = 3\, \frac{\partial G}{\partial z} \Bigg\vert_{z=1}=\frac{3}{2p-1}\,.\end{equation} \\
However, when $p<1/2$, $G(1)=[{p}/({1-p})]^2 <1\,.$ Thus in this case a proportion $1-G(1)$ of the walks do not close; 
they are extended. For such walks, (\ref{eb}) is invalid since it describes only closed walks of a finite length. For $p=1/2$ and 
large 
$L\in \mathbb{Z}$, we find the critical behaviour: $P(e;3L)\sim L^{-\frac{3}{2}}\, .$ The corresponding density of states in the 
quantum system (defined locally on an edge far from the surface, to avoid surface effects) has the behaviour \begin{equation} 
\rho(\epsilon) \propto \begin{cases} \epsilon^2& p>\frac{1}{2}\\ |\epsilon|^\frac{1}{2}& 
p=\frac{1}{2}\\ \text{constant}& p<\frac{1}{2} \end{cases} \end{equation}

We can extend this analysis to consider coordination number $q>3$; then 
\begin{equation}G(z)=z\left[p+(1-p)G(z)\right]^{q-1}\,.\end{equation} Thus 
$z_b=(1-p)^{-1}(q-1)^{1-q}[(q-2)/{p}]^{q-2}$ locates the unique branch point of $G(z)$. 
The complex roots of the two branches of $G(z)$ corresponding to $G(1)=1$ (all paths of finite length) and $G(1)=W$ (where $W$ 
is the fraction of paths which are infinite), coincide at $p=(q-2)/(q-1)$. 
We conclude that for $p>(q-2)/(q-1)$ all walks are localised, whereas for $p<(q-2)/(q-1)$ there exists a fraction $W$ of extended 
walks.
In addition, we find that the critical behaviour $P(e;qL)\sim L^{-\frac{3}{2}}$ holds for all $q\geq3$.
In the localised phase, the typical loop length is given by \begin{equation}<L>=\frac{q}{p-(1-p)(q-2)}\,.\end{equation} Thus, with 
critical value $p_c=1/2$ replaced by $p_c=(q-2)/(q-1)$, the behaviour of the density of states is the same as in the $q=3$ case. The 
critical exponents are therefore universal and independent of $q$. In fact, we note that there is a one-to-one correspondence between 
walks on the Cayley cactus and percolation clusters on the Cayley tree; therefore the critical exponents are related to those of 
percolation.

\section{Discussion}
\label{discussion}

In this section we outline directions for future work. These
are of two kinds. First, one can imagine a variety of models which are likely
to exhibit phenomena not shown in the examples 
treated in Sec.~\ref{applications}. Second, one can hope to
make use of connections between the classical random walks that
arise in our approach and statistical problems studied previously.
Further work on both these aspects is in progress.\cite{manhatten,n-layer}
In general, the equivalent classical problems correspond to self-attractive
random walks of various kinds. They are attractive because the
weight for passing through a given node twice is $p$, rather than $p^2$.
However, the actual behaviour is expected to depend very much on the
individual lattice. Many of these problems correspond to
the classical scattering of light by random arrays of mirrors: for
example, in each decomposition of Fig.~\ref{decomp}, a two-sided mirror can be
placed so that the classical trajectory reflects off it. Such problems
have been studied extensively\cite{gunn-ort,kong,cohen,wang}. They may also be realised in terms of \em
history-dependent \em kinetic random walks, in the sense that the
walker may be thought of as placing a mirror at random the first time
it reaches a node: if it revisits the node, however, the mirror is
already in place. This puts such models in the class of so-called `true'
self-avoiding walks, which have been studied using field-theoretic RG
methods in the limit of weak scattering\cite{obukhov}.
Interestingly, such studies indicate a critical dimension of two, just
as for Anderson localisation.

\begin{figure}
\includegraphics[width=7cm]{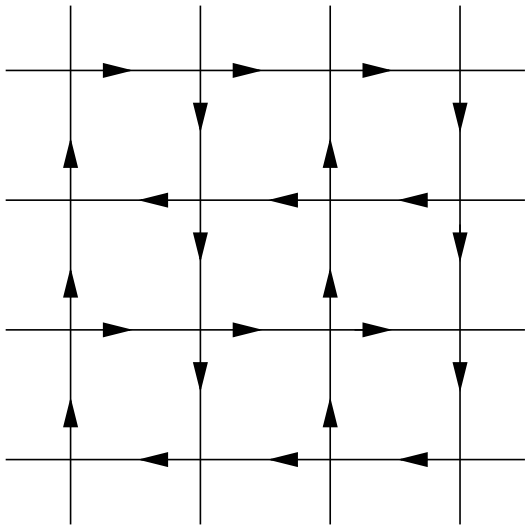}
\caption{The Manhattan lattice. In the network model, $S$-matrix elements
of $\cos\theta$ and $\pm\sin\theta$ are associated respectively
with $90^\circ$ turns at nodes and with propagation in a straight line.}
\label{manhatten}
\end{figure}

A two-dimensional model with behaviour which we expect to contrast with that
found in the spin quantum Hall effect can be obtained by taking $\cal G$
to be the Manhattan lattice, illustrated in Fig.~\ref{manhatten}, in place of
the L-lattice. The crucial distinction is that, for the network model on
the L-lattice, two 
distinct phases can be identified from the different nature of
edge states in the limits $p\to 0$ and $p\to 1$, while for the Manhattan 
lattice there are no edge states in either limit. Equivalently,
the Hall conductance of the model on the L-lattice may be non-zero,
being determined on short distances by the value of $p$, but quantised
at large distances, while on the Manhattan lattice
the Hall conductance always has average value zero. Consequently,
one expects from the scaling flow diagram for 
quantum Hall systems\cite{khmelnitskii} and the renormalisation
group calculations for the class C non-linear sigma model
at weak coupling\cite{sigma} that states of the Manhattan model should be localised
for all $p>0$. Since classical trajectories on the
Manhattan lattice cross at nodes with probability $p$,
they are not the hulls of percolation clusters. Nevertheless,
it is straightforward to use bond percolation on a lattice which
has a lattice spacing larger by a factor $\sqrt 2$ and is rotated
by $45^\circ$ relative to the Manhattan 
lattice to set an upper bound on loop sizes: this is sufficient
to prove localisation for $p>1/2$,
and for $p=1$ classical loops are simply the elementary plaquettes of $\cal G$. 
Conversely, at $p=0$
classical trajectories are simply straight lines, and so one expects the
localisation length to diverge as $p\to 0$. 

A second variant on the spin quantum Hall effect can be constructed
by considering the Sp($N$) model on the L-lattice, but with $N>2$. It is
natural to anticipate from the scaling flow diagram 
proposed for quantum Hall systems\cite{khmelnitskii}
that this model should have $N/2$ delocalisation transitions
as the node parameter $p$ increases from $p=0$ to $p=1$.
Between transitions, states are localised and 
the Hall conductance is quantised. As a transition is approached,
the localisation length diverges, and on passing through a transition,
the number of edge states and the quantised value of the Hall 
conductance both change by two.
It remains a challenge to understand how or whether
such behaviour arises in the language of classical walks,
and might be interesting to relate the large-$N$
limit of the lattice problem to the field theory discussed in 
Ref.~\onlinecite{bernard}
 
A third direction is to consider models defined on graphs in three or more
dimensions. A three-dimensional version of the U(1) network model
has been studied previously using a layered system\cite{chalker-dohmen}
and models with cubic symmetry may also be constructed. For
systems in three and higher dimensions 
one expects that both localised and metallic 
phases should be possible, each existing over a range of values for
node parameter, $p$, with a transition between the two at a
critical value $p=p_c$. Since in two dimensions it is known that the
theta-point transition between collapsed and swollen phases of a
self-attracting polymer chain has the same exponents as those of
percolation hulls\cite{DupSal}, it is tempting to suggest that
in higher dimensions this delocalisation transition might also be in the
theta-point universality class. If so, we would expect mean-field
behaviour, with logarithmic corrections for $d=3$.

\section*{Acknowledgments}
We thank M. Bocquet for a careful reading of the manuscript and M. R. Zirnbauer for valuable correspondence. 
The work was supported in part by the EPSRC under 
Grant GR/J78327.


\begin{thebibliography}{99}
\bibitem{footnote} At this point we
should restrict the $z_j$ all to be real for Lemma 2 to be applicable. 
This does
not affect the argument.
\end{thebibliography}
\end{document}